\RequirePackage{etoolbox}
\patchcmd{\bibliographystyle}{#1}{angew}{}{}

\documentclass[journal=jpclcd,manuscript=article]{achemso}
\usepackage{rsc}
\usepackage[utf8]{inputenc}
\usepackage[T1]{fontenc} % Use modern font encodings
\usepackage[english]{babel}
\usepackage{lmodern}
\usepackage{textcomp}
\usepackage{csquotes}
\usepackage{graphicx}
\usepackage{caption}
\usepackage{wrapfig}
\usepackage{amsmath,bm,amssymb}
\usepackage{chemformula} % Formula subscripts using \ch{}
\setcitestyle{square}
\usepackage{lscape}
\usepackage{setspace}
\usepackage{booktabs}

\author{Carly A. Whittaker}
\email{carly.whittaker2154@gmail.com}
\affiliation[COPL]
{{\normalfont \footnotesize Dr Carly A. Whittaker, Arthur Perret, Charles W. Fortier, Olivier-Michel Tardif, Dr Sébastien A. Lamarre, Steeve Morency, Prof. Younès Messaddeq, Prof. Claudine Nì. Allen}\\ Centre d’optique, photonique et laser (COPL), Département de physique, de génie physique et d’optique, Université Laval, Québec G1V 0A6, Canada}
\alsoaffiliation[CHU PhysMed]{{\normalfont \footnotesize Dr Carly A. Whittaker, Prof. Luc Beaulieu}\\ Département de physique, de génie physique et d’optique, Centre de recherche sur le cancer, CHU de Québec, Université Laval, Québec G1V 0A6, Canada}
\author{Arthur Perret}
\author{Charles W. Fortier}
\author{Olivier-Michel Tardif}
\author{Sébastien A. Lamarre}
\author{Steeve Morency}
\affiliation[COPL]
{{\normalfont \footnotesize Dr Carly A. Whittaker, Arthur Perret, Charles W. Fortier, Olivier-Michel Tardif, Dr Sébastien A. Lamarre, Steeve Morency, Prof. Younès Messaddeq, Prof. Claudine Nì. Allen}\\ Centre d’optique, photonique et laser (COPL), Département de physique, de génie physique et d’optique, Université Laval, Québec G1V 0A6, Canada}
\author{Dominic Larivière}
\affiliation[Radioécologie]{{\normalfont \footnotesize Prof. Dominic Larivière}\\Laboratoire de radioécologie, Département de chimie, Université Laval, Québec G1V 0A6, Canada}
\author{Luc Beaulieu}
\affiliation[CHU PhysMed]{{\normalfont \footnotesize Dr Carly A. Whittaker, Prof. Luc Beaulieu}\\ Département de physique, de génie physique et d’optique, Centre de recherche sur le cancer, CHU de Québec, Université Laval, Québec G1V 0A6, Canada}
\author{Younès Messaddeq}
\author{Claudine Nì. Allen}
\email{claudine.allen@phy.ulaval.ca}
\affiliation[COPL]
{{\normalfont \footnotesize Dr Carly A. Whittaker, Arthur Perret, Charles W. Fortier, Olivier-Michel Tardif, Dr Sébastien A. Lamarre, Steeve Morency, Prof. Younès Messaddeq, Prof. Claudine Nì. Allen}\\ Centre d’optique, photonique et laser (COPL), Département de physique, de génie physique et d’optique, Université Laval, Québec G1V 0A6, Canada}

\title{Colloidal Quantum Dot-Doped Optical Fibers Drawn From Polymer Nanocomposites}


\begin{document}

\begin{tocentry}
%\begin{wrapfigure}{r}{0.5\textwidth}
\includegraphics[height=2cm]{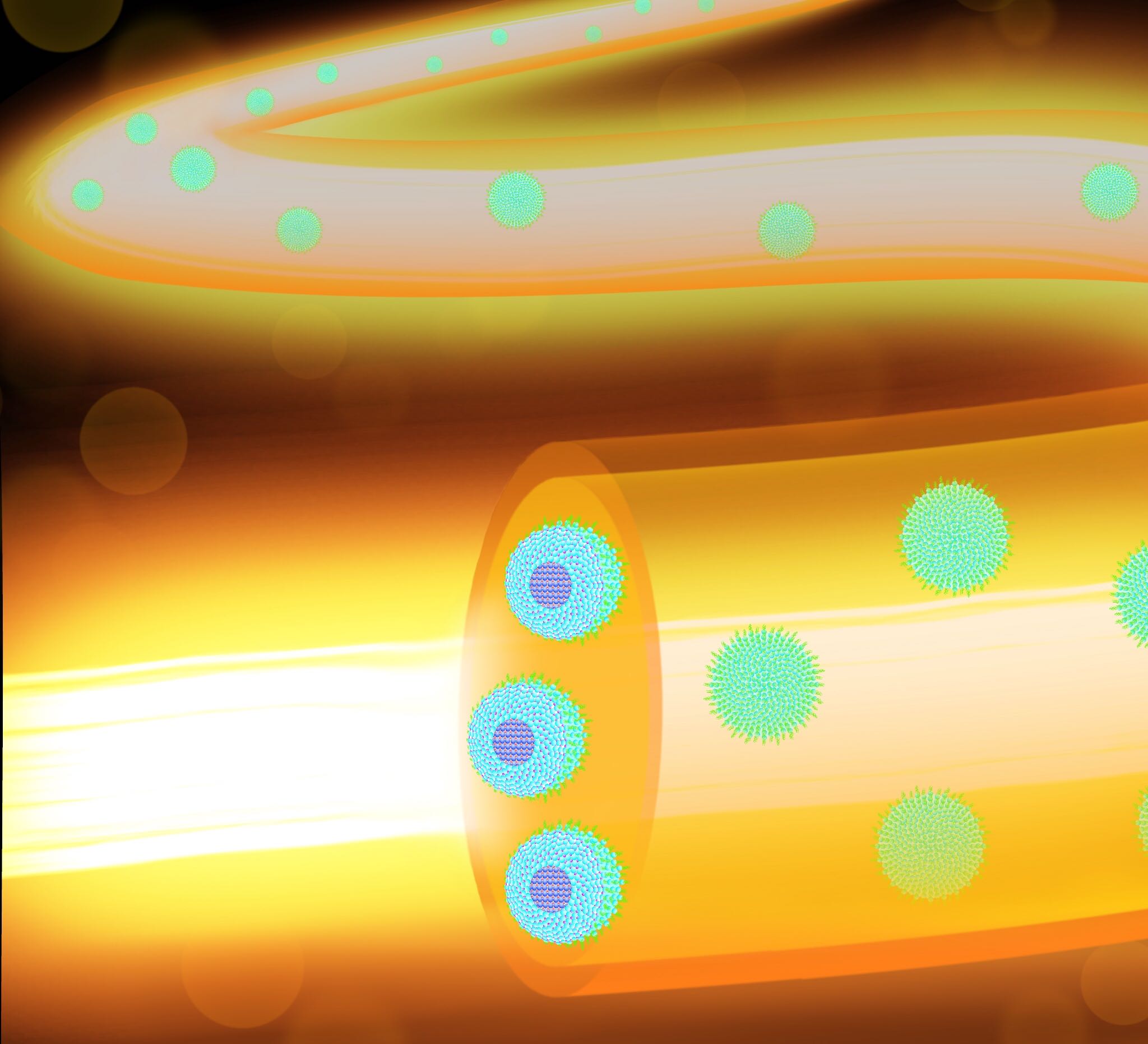} 
Nanocomposite optical fibers with customizable properties are realised through polymerization directly embedding colloidal quantum dots for truly homogeneous doping of the polystyrene core matrix. These hybrid, large area active fibers display a bright and stable luminescence over days of excitation, making them relevant for a wide range of smart applications.
%\end{wrapfigure}
\end{tocentry}

%\setlength{\leftskip}{1cm}
%\begin{minipage}[c]{.9\textwidth}
%\begin{center}
%\setlength{\leftskip}{1cm}

%\footnotesize{
%\noindent {Dr Carly A. Whittaker, Arthur Perret, Charles W. Fortier, Olivier-Michel Tardif, Dr Sébastien A. Lamarre, Steeve Morency, Prof. Younès Messaddeq, Prof. Claudine Nì. Allen}\\ 
%{1. \textit{Centre d’optique, photonique et laser (COPL), Département de physique, de génie physique et d’optique, Université Laval, Québec G1V 0A6, Canada}}\\[1ex]
%\noindent {Dr Carly A. Whittaker, Prof. Luc Beaulieu}\\
%{2. \textit{Département de physique, de génie physique et d’optique, Centre de recherche sur le cancer, CHU de Québec, Université Laval, Québec G1V 0A6, Canada}}\\[1ex]
%\noindent {Prof. Dominic Larivière}\\
%{3. \textit{Laboratoire de radioécologie, Département de chimie, Université Laval, Québec G1V 0A6, Canada}}
%} 
%\end{center}
%\end{minipage}

\begin{abstract}
\begin{onehalfspace}
Colloidal quantum dots (cQDs) are now a mature nanomaterial with optical properties customizable through varying size and composition. However, their use in optical devices is limited as they are not widely available in convenient forms such as optical fibers. With advances in polymerization methods incorporating nanocrystals, nanocomposite materials suitable for processing into high quality hybrid active fibers can be achieved. We demonstrate a plastic optical fiber fabrication method which ensures homogeneous dispersion of cQDs within a polymer core matrix. Loading concentrations between 10$^{5}$-10$^{7}$ CdSe/CdS cQDs per cm$^{3}$ in polystyrene were electronically imaged, confirming only sporadic sub-wavelength aggregates. Rayleigh scattering losses are therefore dominant at energies below the semiconductors’ band gap, but are overtaken by a sharp CdS-related absorption onset around 525~nm facilitating cQD excitation. The redshifted photoluminescence emission is then minimally reabsorbed along the fiber with a spectrum barely affected by the polymerization and a quantum yield staying at $\sim$65$\%$ of its initial value. The latter, along with the glass transition temperature and refractive index, is independent of the cQD concentration hence yielding a proportionally increasing light output. Our cQD-doped fibers are photostable to within 5$\%$ over days showing great promise for functional material applications.
\end{onehalfspace}
\end{abstract}

\section{Introduction}
Optical fibers are now commonly known as the Internet backbone for light transmission, whereas nanomaterials are a motor for customizing matter properties for innovative applications \cite{Bourzac,Bio,Bera,chen2}, thus the meeting of both technologies in hybrid nanocomposite optical fibers forms part of an entirely new research direction\cite{hybrid}. Such development is geared towards one-dimensional and flexible photonic devices with an increasing number of functionalities easily integrable in diversified applications, most notably smart and wearable ones. For example, large area active fibers are already used in imaging \cite{Abouraddy, sorin}, remote and distributed sensing \cite{gumennik}, dosimetry\cite{TP, Delage3}, and functional fabrics \cite{Sorin2}. Expanding functionalities further can harness the broad appeal of colloidal quantum dots (cQDs) that stems largely from their unique optical properties, namely the tuneability of their narrow photoluminescence (PL) emission spectrum with respect to semiconductor nanocrystal size, shape and composition\cite{Peng,Burda} as well as their wide excitation spectrum. Nonetheless, cQDs are seldom used as a stand-alone optical material on account of their lack of environmental stability and user friendliness. The most straightforward and robust means of exciting and collecting the light produced by cQDs is to incorporate them directly within the core material of an optical fiber, thereby mutually leveraging hybrid functionality. There in the core, these semiconductor nanocrystals can interact maximally with the guided electromagnetic waves, transporting the emitted radiation to the desired location, while remaining mostly protected from the outside environment.\par
Historically, one could argue the first quantum-confining nanocrystals were made by melting both glass and semiconductor precursors at high temperature\cite{Ekimov,Borrelli87,Mauro,McMillan,Borrelli94}, nucleating quantum dots throughout the glass structure, hence naturally yielding preforms for the first category of nanocomposite optical fibers\cite{dong,Liu,Bzheumikhov,Bhardwaj2,Peng19}. This glass melt-quenching method does not allow for elaborate control of the nanocrystal surface passivation\cite{Xia,Lee19}, thus defects degrade the PL emission, with further challenges to control the density of nanocrystals and their size distribution\cite{Jacob,Bhardwaj,Huang,Huang2,Xia19,Wei19}. A second category of debatably nanocomposite fibers thus recently emerged doping optical fibers with cQDs using a number of post-processing techniques, such as filling hollow core fibers with a liquid containing cQDs\cite{Zhang,Hreibi,Cheng} and coating the core of microstructured optical fibers\cite{meissner,Yu3,dehaven,Bravo} or fiber tapers and ends\cite{bastida,Sun,Delage2}. These techniques seem \textit{a priori} advantageous in their chemical simplicity and flexibility, benefiting from the separate synthesis of cQDs under well controlled conditions enabling passivation with heterostructures of multiple semiconductor shells\cite{reiss,book} and even with further surface ligand exchange afterwards\cite{Owen,Pong,Burda}. However, the lack of stability and robustness coupled with high optical losses in the cQD host solvents limits the practicality of liquid core fibers, with the added drawback of air exposure when the solvent evaporates, causing surface chemical reactions that increase photobleaching and photoactivation.\par
A third nanocomposite fiber category is thus sought to achieve a homogeneous dispersion of cQDs in a solid host material, but, along with minimizing aggregation and surface damage, this remains one of the greatest challenges to cQD doping techniques even in the bulk\cite{agg1,Nowaczynski,Wei19}. Nanocomposite fiber fabrication with cQDs was attempted using either glass or polymer host matrices. In the former case based on a modified chemical vapor deposition (MCVD) process at high temperature, PL emission was not reported and absorption spectra hint that quantum confinement of the unprotected CdSe cores was completely lost\cite{Ju,Watekar,Watekar2}. In the latter case, polymer mixtures with pre-synthesized cQDs are either electrospun\cite{Liu2,Li,Zhu,Kumar} or directly drawn\cite{Yu,Meng,Yang} while the solvent rapidly evaporates at room temperature. The emission spectra of the cQDs within the exposed fiber cores usually seem to agree well with those in solution, however the limitation in these methods lies in the processing itself. Indeed, no protective outer cladding can be spun or drawn this way, thereby severely hindering the tenuous waveguided light propagation as shown by the PL irradiance output changing with excitation position along the fiber according to the supporting substrate\cite{Yu}. These fabrication methods also overlook the established thermal drawing know-how, whereby complex preforms are melted in a furnace to enable optimization and innovation in optical fiber design\cite{large}. In this category, quantitative optical characterizations are so far sparse and no clear proof of uniform cQD doping has been provided, even for the most promising strategies starting to overcome the lack of cladding, namely a mixture of cQDs and photocurable adhesive in a hollow fiber for laser applications\cite{Cheng2,Cheng17,Cheng18} or a fused rod made of poly(methyl methacrylate) (PMMA) and cQDs sleeved into a PMMA tube and microstructured fiber\cite{yu2,Barton}.\par
Here, we report the first plastic optical fiber truly doped with colloidal semiconductor nanocrystals, namely cQD emitters, with robust, bright and stable luminescence. Unlike the methods mentioned above, our approach allows the cQD dopants to be controllably and homogeneously dispersed within the fiber core material with minimal impact on their luminescence properties. In-house synthesized cQDs are dispersed in a liquid monomer, in this case styrene, to polymerize a nanocomposite preform, which is subsequently drawn into fiber using the direct thermal method. The addition of well-passivated CdSe/CdS cQDs to the shielding host matrix makes use of the flexibility provided by a separate hot-injection colloidal synthesis while the resulting polymer nanocomposite enables the simple, low-temperature direct drawing of clear and highly photostable optical fibers. In the first part of this paper, we discuss the design and fabrication of our cQD-doped fibers, optimizing their composition and geometry for large area active fibers applications. Extensive characterization of nanocomposite preforms and derived optical fibers with varied cQD dopant concentrations has been performed, with physical characterization measurements of the glass transition temperatures $T_g$ and cQD spatial distributions in the second part, followed by optical characterization results ranging from emission and extinction spectroscopies to refractometry in the last part.

\section{Results and discussion}
 \subsection{Material selection and fiber design}
 
Polystyrene was selected as the cQD host matrix for the core material of the nanocomposite fiber on account of its high refractive index, sufficient transparency in the cQD emission region, and the ease at which it can be thermally processed. The low softening temperature of polystyrene also ensures minimal degradation of the cQDs during the fiber drawing process. These cQD dopants, acting as active light emitters, have a well-known semiconductor heterostructure, namely a CdSe core passivated by a gradient towards a nearly lattice-matched CdS shell\cite{nasilowski,Mahler,Chen,Dai,Lachance-Quirion}. Stress is limited in such nanocrystals, further relaxed with ongoing annealing from the continuous heating during shell growth, thus minimizing crystal defect formation.\\
All cQD doped optical fibers have been fabricated with a PMMA outer cladding. The lower refractive index of PMMA (n$\sim$1.49 @ 650 nm) compared to polystyrene (n$\sim$1.59 @ 650 nm) provides a refractive index contrast helpful to guide light within the fiber core, and also acts as a protective coating against the surrounding environment. The relatively large refractive index difference between polystyrene and PMMA also ensures a large numerical aperture of NA$\sim$0.58 to efficiently collect the light produced by the cQDs in the fiber core. We have fabricated fibers with outer diameters of 1000~$\mu$m $\pm$ 10$\%$ (around 820~$\mu$m core diameter and 90~$\mu$m cladding thickness), resulting in an active volume of around 67$\%$. 

\subsection{Fabrication of cQD-doped optical fibers}

cQD doped optical fibers are fabricated using a two-step process involving the creation of a core preform which is subsequently cladded, annealed, then drawn into an optical fiber. The core preform is made by dispersing a known volume of cQDs in a liquid monomer (styrene) and initiating the polymerization process to embed the nanocrystals within the polymer network. Specifically, our nanocomposite preforms were fabricated with core/shell CdSe/CdS cQDs using volumes of up to 80~$\mu$L from a $\sim$6~$\mu$M stock solution\cite{nasilowski}. Throughout the article, the initial aliquot volume of cQDs, added to a fixed 10 mL nominal volume of styrene will be used to refer to the approximate concentration of cQDs in the core material of the optical fiber. Taking into account the $\sim$15\% polymerization shrinkage of styrene, these aliquot volumes correspond to an approximate concentration of cQDs between 10$^{5}$ to 10$^{7}$~cQDs/cm$^{3}$ in the bulk polymer. Core preforms fabricated with varied cQD volumes are used to explore the effect of dopant concentration on $T_{g}$, cQD spatial distribution, and various optical properties of the nanocomposites. Additionally, a blank polystyrene preform without cQDs has been polymerized under identical conditions for use as a reference sample. \\
The thermal and mechanical properties of polymers are influenced by a range of factors, including their molecular weight, degree of cross-linking between polymer chains and plasticiser content (impurities, cQDs, etc.), all of which will govern their behaviour during fiber drawing. In particular, the molecular weight, related to the average length of polymer chains will have a profound influence on the materials’ processing temperature: on average, longer polymer chains will suffer a greater degree of entanglement than shorter chains, making them more elastic and resistant to flow. Thus, higher molecular weight polymers will require higher temperatures to overcome this elasticity and flow into fiber. Polymerization conditions such as temperature, initiator concentration and cQD concentration are controlled to produce a polymer with a molecular weight distribution which allows the preform to be drawn smoothly at a suitably low temperature still above the corresponding $T_{g}$\cite{large}. The nanocomposite core preforms produced using this technique appear transparent and free of visible light scattering, even at the highest dopant concentrations used, as can be seen in Figure~1(a).\\
Each nanocomposite preform is then stacked within a commercial PMMA tube and the two materials are heated above their $T_{g}$, and co-drawn into an optical fiber. It is important that both the core and cladding polymers possess similar thermal properties in order for them to soften and flow evenly at a similar temperature, around 160$^{\circ}$C. As such, the key parameter of initiator concentration used in the polymerization of the core preform was selected to not only ensure it draws at a suitable temperature, but that it also possesses a similar $T_{g}$ to the PMMA cladding. A cross-section micrograph of a typical fiber obtained is shown in Figure~1(b), and in 1(c), an image showing the PL produced in a $\sim$1.5~m length of fiber excited with an ultraviolet (UV) lamp.

\begin{figure}[htpb!]
\centering
\includegraphics[height=8cm]{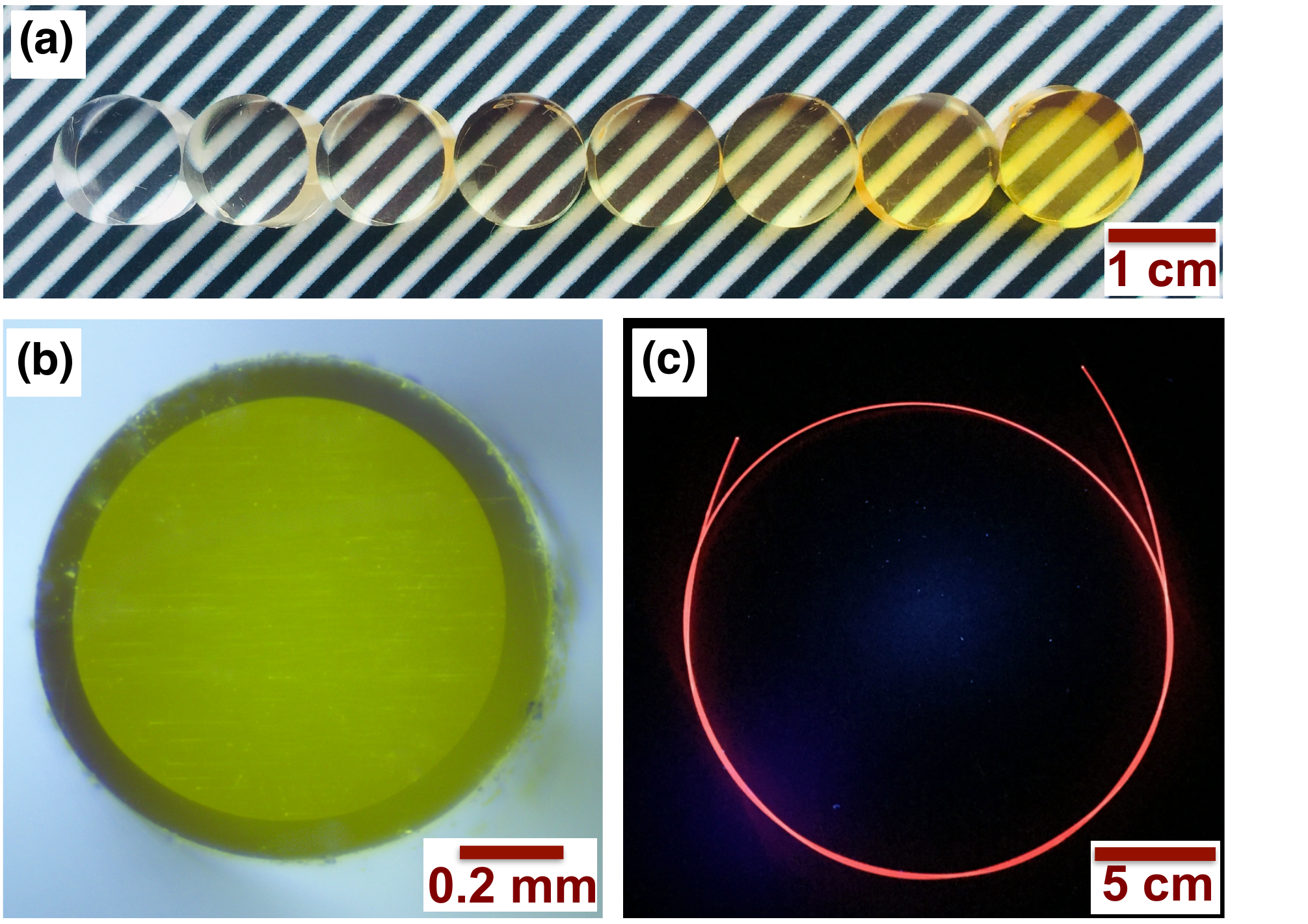} 
\caption{(a) Polystyrene preforms with cQD dopant concentrations ranging from 1~$\mu$L (far left) to 80~$\mu$L (far right) per 10 mL styrene. (b) Micrograph of the endface of our nanocomposite optical fiber showing the cQD-doped polystyrene core of ~820~$\mu$m in diameter and the ~90-$\mu$m thick PMMA cladding. (c) Photograph showing the PL emitted by our cQD doped fiber under a UV lamp.} \label{fig:pretty}
\end{figure}

\subsection{Glass transition temperature}

The $T_{g}$ values, calculated as the onset temperature of the baseline shift in the differential scanning calorimetry (DSC) thermogram of each polymer, are listed in Table 1. The blank polystyrene was found to possess a $T_{g}$ of (100$\pm$2) $^{\circ}$C, similar to that measured for the PMMA cladding material (97$\pm$2)$^{\circ}$C. One particular concern when adding dopants to a polymer matrix is the decrease in $T_{g}$ due to the plasticisation effect \cite{koike}, whereby the addition of a dopant introduces free volume into the polymer which can promote polymer-diluent interactions in place of polymer-polymer interactions. Additionally, an increase in heat transport within the polymer matrix can occur when semiconductor materials, such as cQDs, are incorporated within the polymer matrix, which can also influence the $T_g$ of the composite material \cite{saldivar}. In our case, as a result of the low dopant concentrations used, all less than 10$^{7}$~cQDs/cm$^{3}$, the magnitude of these effects is small, with at most, a 1$^{\circ}$C decrease in $T_{g}$ over the whole dopant concentration range.

\vspace{1cm}
\renewcommand{\arraystretch}{1.5}
\begin{landscape}
\begin{table}[htpb!]
\centering
\caption{Properties of fibers fabricated with the listed nominal volume of a 6~$\mu$M cQD stock solution in 10 mL styrene to vary the cQD core dopant concentration: glass transition temperature $T_g$, mean separation distance between cQD nearest neighbours in the preforms, percentage cQD pair clusters with null separation and optical losses of light transmitted through the fiber $\alpha$. N/A = not applicable, N/M = not measurable, data out of analysis or instrument dynamic range
}

\begin{tabular}{@{}ccccccc@{}}
\toprule
\textbf{\begin{tabular}[c]{@{}c@{}} \textbf{cQD Concentration} \\ \\ {}\bm{$\mu$}L/10 ml styrene{}\end{tabular}} & \textbf{\begin{tabular}[c]{@{}c@{}}\bm{$T_{g}$} \\ \\ {}$^{\circ}$ C{ $\pm$ 2$^{\circ}$C}\end{tabular}} & \textbf{\begin{tabular}[c]{@{}c@{}}Neighbouring \\ cQD distance \\   {}nm{}\end{tabular}} &   \textbf{\begin{tabular}[c]{@{}c@{}}Proportion of\\ cQD pair clusters \\  \vspace{0.2cm} \%\\ \end{tabular}} & \textbf{\begin{tabular}[c]{@{}c@{}}\bm{$\alpha_{abs}$}(525 nm)\\ \\ {}dB/m{}\end{tabular}} & \textbf{\begin{tabular}[c]{@{}c@{}}\bm{$\alpha_{sc}$}(650 nm)\\ \\ {}dB/m{}\end{tabular}} & \textbf{\begin{tabular}[c]{@{}c@{}}\bm{$\alpha_{sc}$}(800 nm) \\ \\ {}dB/m{}\end{tabular}} \\ \midrule
0 & 100 & N/A & N/A & 6.6 $\pm$ 0.4 &4.2 $\pm$ 0.3 & 3.1 $\pm$ 0.3  \\
1 & 100 & N/M & N/M & 7.1 $\pm$ 0.3  & 4.3 $\pm$ 0.3 & 3.4 $\pm$ 0.3 \\
5 & 99 & N/M & N/M & 7.3 $\pm$ 0.3 &4.3 $\pm$ 0.3 & 4.1 $\pm$ 0.3  \\
10 & 100 & N/M & N/M & 10.1 $\pm$ 0.4 &4.3 $\pm$ 0.5 & 3.7 $\pm$ 0.5  \\
20 & 99 & N/M & N/M & 13.8 $\pm$ 0.5 &4.7 $\pm$ 0.6 & 3.4 $\pm$ 0.6  \\
30 & 100 & 912 & 4 & 19.3 $\pm$ 0.7 &5.5 $\pm$ 0.3 & 4.0 $\pm$ 0.3  \\
40 & 99 & 500 & 14 & 29.7 $\pm$ 0.4 &8.0 $\pm$ 0.4 & 4.6 $\pm$0.4  \\
60 & 99 & 468 & 8 & N/M &11.0 $\pm$ 0.6 & 5.4 $\pm$ 0.9  \\
80 & 99 & 453 & 2 & N/M  &17.0 $\pm$ 1.1 & 6.8 $\pm$ 1.1 \\
PMMA & 97 & N/A & N/A & N/A & N/A & N/A \\ \bottomrule
\end{tabular} \label{tab:tab}
\end{table}
\end{landscape}

\subsection{Transmission electron microscopy of nanocomposite preforms} \label{ss:tem}
Transmission electron microscopy (TEM) micrographs of the cQDs in hexane and preform are shown using different scales in Figures~2(a) and (b). The particle size distribution of cQDs in polymer presented in Figure~2(c) follows a log-normal distribution \cite{goodwin} with a mean diameter of 15.4~nm and is relatively monodispersed with a variance of $\sim$6.2~nm$^{2}$. This is similar to the distribution in hexane (see Figure~\ref{fig:sizesupp} in supporting info.) displaying a mean diameter of 14.1~nm, which indicates the polymerization process has little to no effect on cQD size. 

The separation between the surfaces of neighbouring cQDs in polystyrene for dopant concentrations between 30~$\mu$L to 80~$\mu$L was calculated by measuring the distance between their centroids in the polystyrene matrix and subtracting both radii. The cumulative cQD distance distributions for each dopant concentration are displayed in Figure~2(d)-(g), showing the proportion of analysed cQDs which have a nearest neighbour within a given distance. Polymers with low dopant concentrations, 20~$\mu$L of cQD solution (always in 10 mL styrene) and below, were not included in the analysis as the separation between neighbouring cQDs was generally too large to fit within an image frame at sufficient resolution. As the cQD concentration in polymer increases from 30~$\mu$L to 80~$\mu$L, the mean neighbouring cQD distance decreases as expected (Table~1), and the proportion of cQDs with a nearest neighbour closer than 1~$\mu$m rises by about 30$\%$ from ~65$\%$. \\
The vast majority of cQDs observed in the polymer samples appeared well dispersed, however small clusters were sporadically observed as exampled in Figure~\ref{fig:temsupp} of the supporting info. The above nearest neighbour analysis did not show a clear correlation in Table~1 between the cQD concentration in polymer and the small proportion of observed cQD pair clusters. If these clusters occurred entirely randomly, their spatial distribution should follow a Bernoulli process, but the cumulative distributions at short distances in Figure~2(d)-(g) exceed significantly the statistical model and do not decrease asymptotically as seen in Figure~\ref{fig:cQDs} of the supporting info. The clustering therefore results from aggregation tentatively attributed to a depletion attraction between cQDs during the growth of polymer chains \cite{Hu, hooper}. Longer oligomers may indeed be excluded from gaps between cQDs wandering nearby in the remaining monomer solvent phase. Aggregation then arises from the osmotic pressure through the effective inward attraction as cQDs pass each other.\\
\newpage
To hamper depletion attraction, our preforms were prepared using hot sonication of the cQD-doped monomer mixtures in a heated water bath for 7h. Preforms fabricated without this pre-polymerization step display large aggregates of cQDs, with little to no evidence of isolated cQDs found in any polymer samples. TEM images of nanocomposites prepared with and without hot sonication are presented in the supporting info. in Figures~\ref{fig:temsupp} and~\ref{fig:aggsupp}(a) respectively. This sonication step therefore minimizes optical scattering, ensuring the necessary transparent quality for optical fibers that will be further investigated in the following sections. 

\begin{figure}[htpb!]
\centering
\includegraphics[height=12cm]{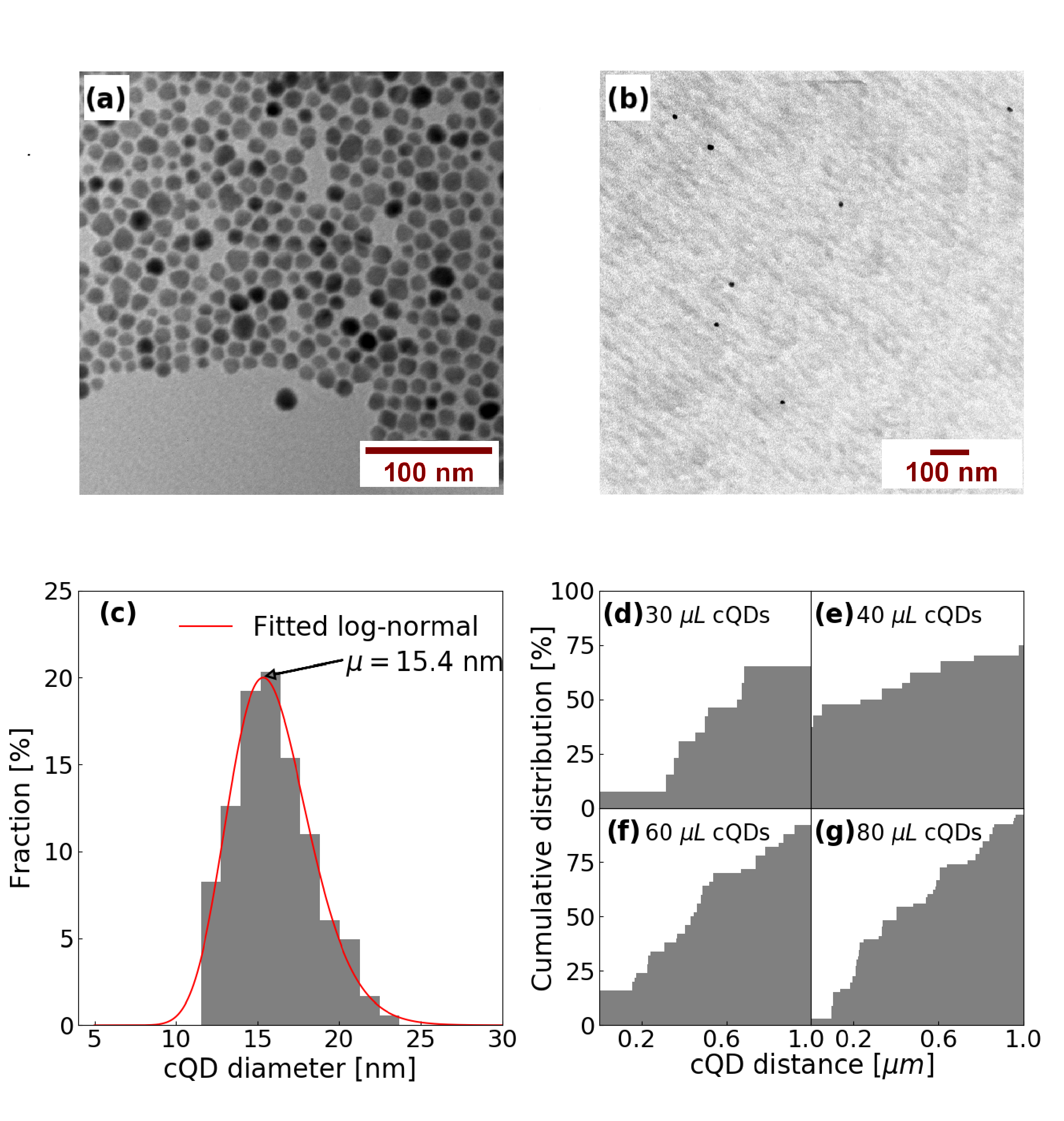} 
\caption{TEM micrographs of cQDs dispersed in (a) hexane and (b) polystyrene preform at the highest cQD concentration. (c) Size distribution of cQDs in polystyrene as measured from TEM data of all doped samples, fitted to a log-normal probability density (red line). (d) - (e) Nearest neighbour statistics with cumulative histograms of inter-cQD distances in polystyrene for 30~$\mu$L to 80~$\mu$L cQD concentrations. Only cQD pairs residing within the same image frame can be counted and lone cQDs complete each sample distribution.} \label{fig:tem}
\end{figure}

\subsection{Photoluminescence and absorption spectroscopy}

To assess the impact on the cQD optical properties as a result of the fiber fabrication process, the PL and absorption spectra of the cQDs have been measured in polystyrene preform and optical fiber then compared to a reference solution of cQDs in hexane, all shown in Figure~3(a). In order to highlight general absorption features seen in all cases, the light transmission data was processed to remove the host matrix contributions and converted to the same linear optical power scale $P_{in}/P_{out}$ after normalization to the cQD-free reference spectra. In addition to the onset of CdS absorption around 525~nm, a second broad absorption feature may be observed around the 600-nm range at higher cQD concentrations, which in accordance with Nasilowski \textit{et al.}, can be attributed to the CdSe core. Although the core comprises only a small fraction of the total cQD volume, it has a band gap triggering absorption above $\sim$1.75~eV ($<$700~nm) and it can interdiffuse with the CdS shell to form a CdSSe gradient.\\
Regarding the PL spectra, no appreciable change in the cQD emission is observed between solution, preform and fiber samples. This indicates the nanocrystal structure and morphology were not significantly affected by the fiber fabrication process. Following polymerization, the cQDs display a slightly wider (+3~nm FWHM) and skewed PL spectrum, with no further changes evident following preform heating and fiber drawing. This larger spectrum was found to redshift by 4~nm for the maximum, but unevenly on the left and right side by 2~nm and 5~nm respectively. Assuming differences in absorption between polystyrene and hexane are not significant enough to create a filtering distortion on the cQD emission range, two plausible mechanisms could account for the minor changes in PL of the nanocomposite material: cQD aggregation and surface chemical reactions during the polymerization procedure. In the former case, aggregates of cQDs are able to redshift the PL via Förster resonance energy transfer (FRET) \cite{chou}, but this process is likely hindered by the thick CdS shell \cite{Pal}. Results in Section~3.4 firstly show that aggregates represent only a small fraction of all cQDs in polymer, hence changes in the PL spectra are unlikely to originate from inter-dot dipole-dipole coupling inducing energy transfer from smaller to bigger clustered cQDs. Secondly, the PL emission spectrum in polystyrene is identical between samples containing aggregated versus well dispersed cQDs as verified in Figure \ref{fig:aggsupp}(b) of the supporting info. Therefore, the redshift more likely originates from the latter mechanism affecting the cQD surface passivation during thermally activated polymerization, resulting in shallow trap states emitting at longer wavelengths\cite{Chen05,Giansante,Rabouw,Burda}. Additional non-radiative channels reducing the PL emission are also expected from these surface defects, an effect further investigated in section 3.8.\\
Normalised PL spectra of preforms at selected cQD concentrations are shown in Figure~3(b), with the fractional deviation in PL intensity from the 5~$\mu$L sample in Figure~3(c) for all preforms within the 5~$\mu$L to 80~$\mu$L dopant range. The 1~$\mu$L sample data is not included as it was below the noise floor of the instrument. All these PL spectra are identical within 5$\%$ variation at any given wavelength, indicating the cQD dopant concentration in this range has little to no effect on their PL emission spectrum.

\begin{figure}[htpb]
\centering
\includegraphics[height=8cm]{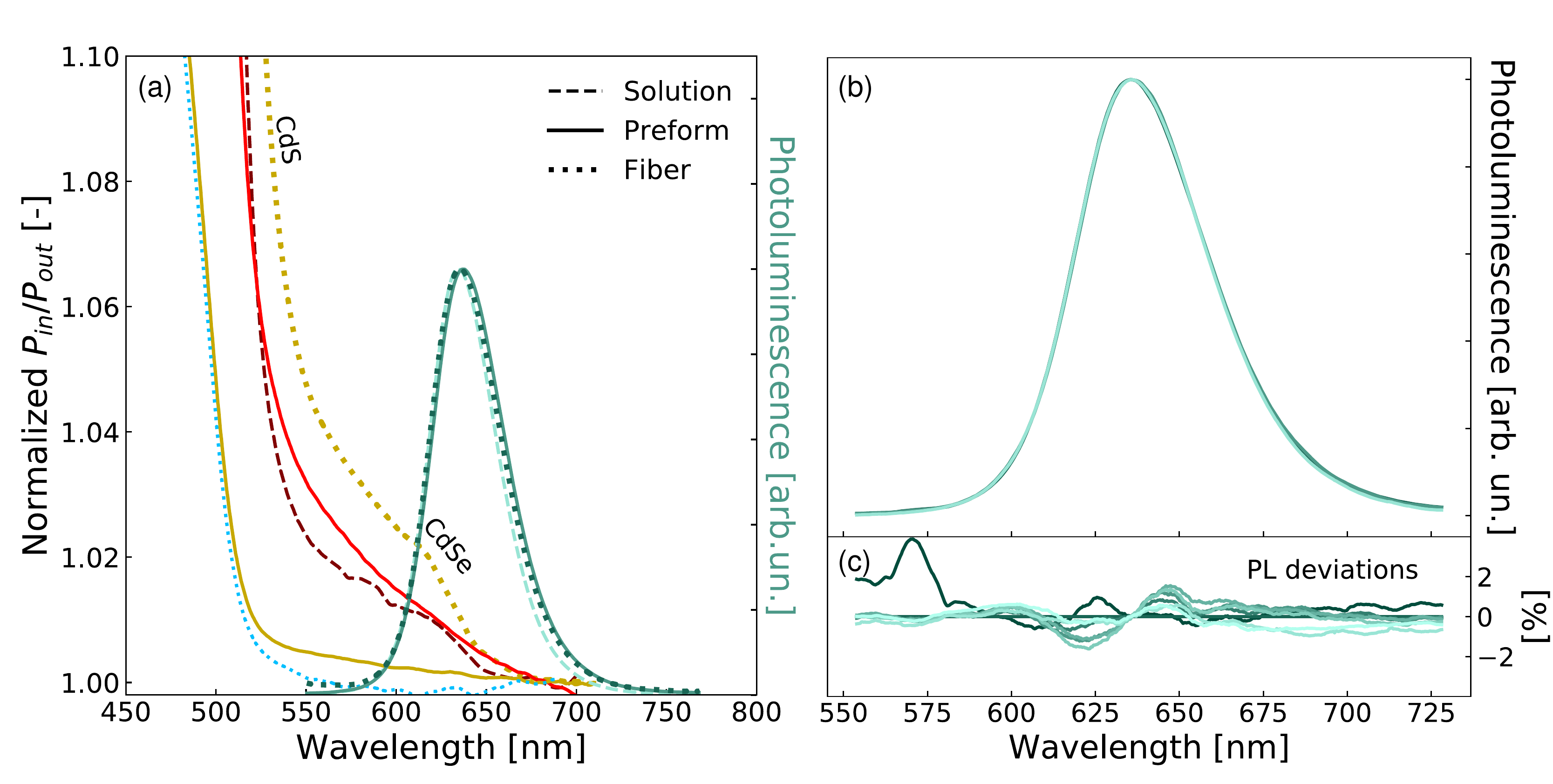} 
\caption{Optical spectroscopy of the nanocomposites. (a) Absorption features of cQDs in the hexane reference solvent (brown dashes), preform (20~$\mu$L yellow line, 40~$\mu$L red line) and fiber (1~$\mu$L cyan dots, 20~$\mu$L yellow dots).  Normalized PL spectra are also shown for 20~$\mu$L cQD samples in solution, preform and fiber. All spectra are similar, nearly independent of the host matrix and cQD concentration as shown in (b): the graphic displays the PL of 5~$\mu$L, 40~$\mu$L and 80~$\mu$L doped preforms, shaded with progressively lighter green and normalized relative to their peak PL intensity. Any deviation stays below 5$\%$ as quantitatively seen in (c) relative to that of the normalized PL intensity of the 5$\mu$L sample.} \label{fig:abspl}
\end{figure}

\subsection{Transmission of cQD-doped fibers}

The overall transmission loss of an active optical fiber will influence a host of performance related properties including its maximum practical length, doping emitter spectrum and collected light irradiance output. The attenuation of each fiber is presented in Figure~4, on a wavelength range overlapping with the cQD spectra previously shown. In this region, many mechanisms will contribute to light attenuation in plastic optical fibers, including absorption and scattering losses \cite{loss}. Absorption effects arise from the active cQD dopants, impurities in the fiber core and from the harmonics of molecular vibrations and stretching of carbon-hydrogen bonds in polystyrene. Scattering from the cQDs and from structural imperfections in the fiber (\textit{e.g.} dust, bubbles, scratches, core diameter variations) can also contribute to fiber attenuation.

\begin{figure}[htpb]
\centering
\includegraphics[height=9cm]{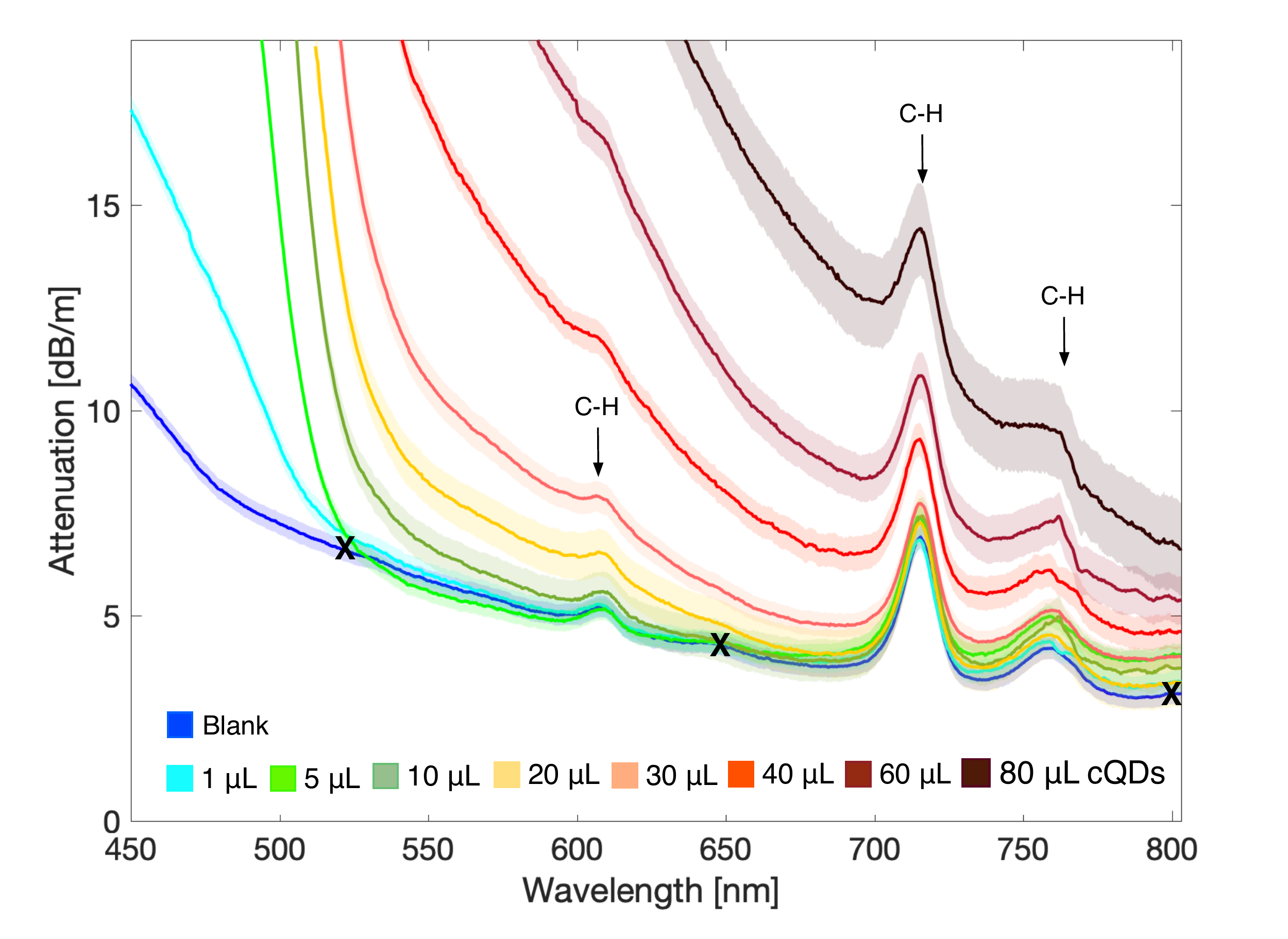} \caption{ Attenuation of the nanocomposite fibers with both cQD and host matrix contributions.  Absorption peaks corresponding to C-H stretching vibrations in polystyrene (arrows) \cite{ch}, $x$ marks for the wavelengths where losses in Table~1 are taken, and measurement uncertainties (shaded regions) and are shown.} \label{fig:loss}
\end{figure}

Selected fiber transmission losses are showcased in Table~1 at wavelengths representative of dominant absorption losses $\alpha_{abs}$, namely 525~nm, and dominant scattering losses $\alpha_{sc}$, namely 650~nm and 800~nm. All doped fibers display a rapid increase in attenuation at wavelengths near 525~nm attributed to the onset of absorption mentioned in the previous section above the CdS band gap energy, with $\alpha_{abs}$ rising steadily with increasing cQD concentration as expected. These transmission losses in the fiber, although considerably large, are indicative of a high absorption efficiency of the active dopants, which is necessary for the production of PL in the nanocomposite material. \\
On account of the small cQD size being inferior to the visible wavelengths of interest, we assume Rayleigh scattering to be the dominant contribution to $\alpha_{sc}$ caused by the cQDs and their aggregates in doped fibers. As the magnitude of this scattering effect varies considerably with wavelength ($\propto\lambda^4$), the attenuation values at 640~nm and 800~nm are compared since these wavelengths do not coincide with any C-H absorption peaks and are located far from the major CdS absorption feature. With low doping levels of 1~$\mu$L and 5~$\mu$L, fiber losses beyond 525~nm appear to lie within the range measured for the blank reference fiber, indicating scattering caused by the cQDs is too low to have a measurable effect on the fiber attenuation. As the cQD concentration is further increased, there is little variation in scattering losses $\alpha_{sc}$($\lambda$ = 800~nm) measured below the CdSe and CdS band gaps, with at most a 4~dB/m difference across the range of doped fibers. This result agrees with the diminishing contribution of Rayleigh scattering and absence of cQD absorption at this wavelength. In contrast, losses $\alpha_{sc}$($\lambda$=650~nm) above the CdSe band gap increase rapidly with dopant concentration, with a $\sim$13~dB/m higher loss observed between the 80~$\mu$L and 1~$\mu$L doped fibers. 

\subsection{Photoluminescence stability and linearity}

The stability of light-emitting fibers is crucial for any application that is not meant to be disposable. Whereas most air-exposed core/shell cQDs are known to photobleach on a timescale of hours or less \cite{Sark,Qin}, Figure~5(a) shows an outstanding PL resistance of our nanocomposite fibers under continuous laser excitation ($\sim$400~nm, $\sim$75~$\mu$W) over 12 days. The integrated PL peak is monitored over time in Figure~5(c). Deviations of less than 5\% from the initial PL intensity are observed as a result of photobrightening and photodarkening effects, with a 5-day period where the stability remains within 1\%. Although the exact activation and bleaching mechanisms are not fully understood, two likely causes are a photoactivated annealing effect passivating a number of nanocrystal defects, or a photoactivated chemical reaction affecting the cQD surface \cite{bright2,bright,Sark}. As some of these processes may be reversible, the illumination source was removed from the fiber after 7 days for a period of 19 hours to check if any signal degradation was recoverable. A slight variation in PL intensity can be observed following the replacement of the illumination source, but no significant recovery is apparent. Overall, after the cQD surface is perturbed during polymerization as will be further evidenced in the next section, the resulting passivation and mostly oxygen-free environment yields a remarkably stable PL emission.\\
The expected linearity of the PL irradiance with cQD concentration was verified for 5-mm fiber segments, and is shown in Figure~5(b). The signal from the 1~$\mu$L doped fiber was within the instrumental noise floor, which is indicated in red. The PL intensity increases proportionally with dopant concentration, with a maximum 2\% deviation from the fitted linear trend (Figure~5(d)). There is thus no evidence of saturation of the PL signal in this concentration range and excitation conditions, with minimal reabsorption along the fiber thanks to the large shift between PL emission and CdS absorption onset \cite{Meinardi, krumer, Coropceanu}. This is consistent with the quantum yield (QY) reported as independent from cQD concentration in the following section, the light irradiance at the fiber output is then proportionally set by the total number of emitters per unit volume. A further increase in scattering loss with concentration and other deviations from the Beer-Lambert law, along with any change in QY, reabsorption and excitation conditions, would eventually be expected to affect the linear behaviour observed here. These results demonstrate the ability to vary the PL intensity of the optical fiber, hence its signal sensitivity for sensing applications by changing the dopant concentration within the fiber core.

\begin{figure}[htpb]
\centering
\includegraphics[height=7cm]{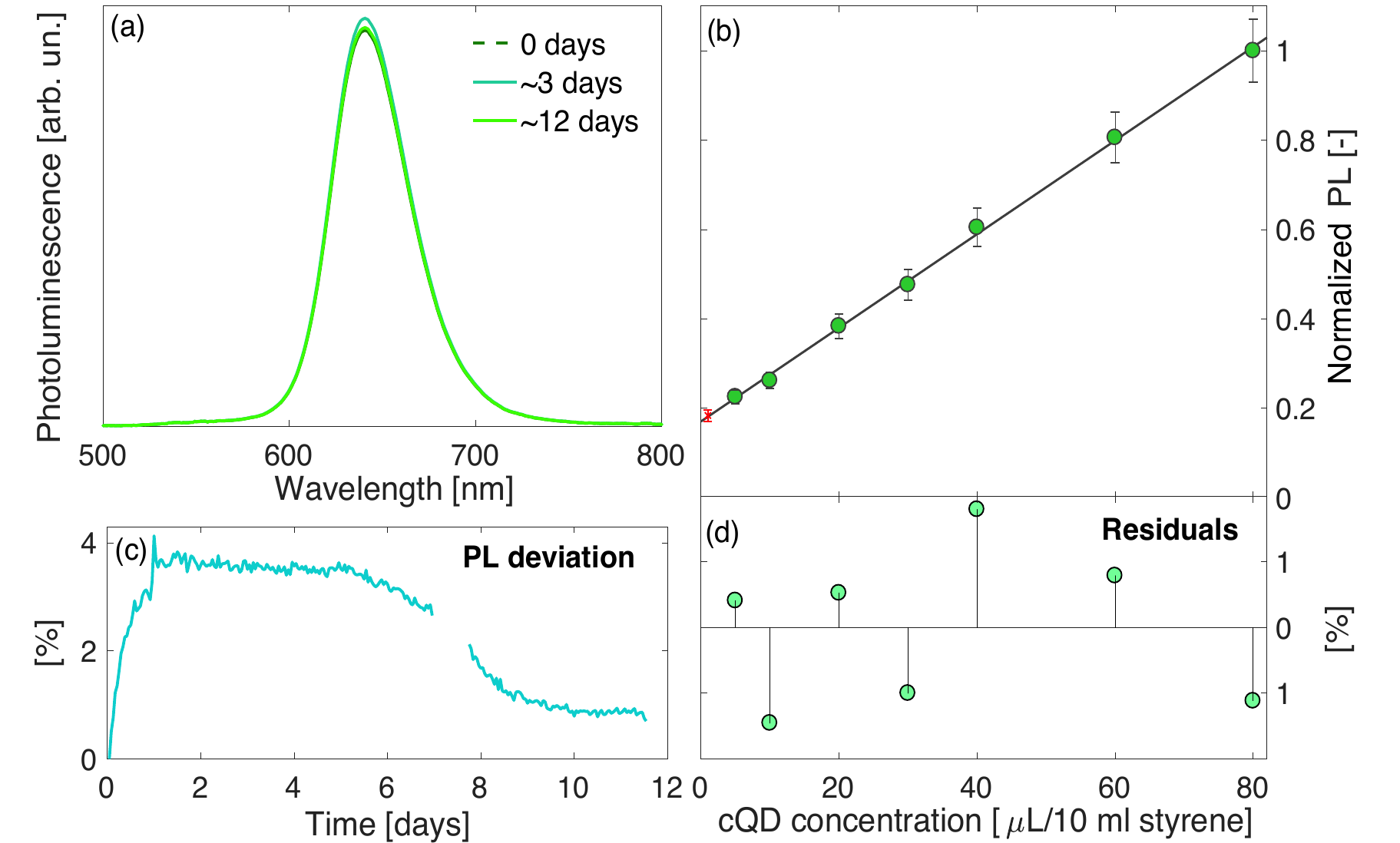} 
\caption{(a) Stable PL spectra produced by a 20~$\mu$L cQD-doped fiber after 0, $\sim$3 and $\sim$12 days of continuous illumination. (b) Integrated PL intensity as a function of cQD dopant concentration in the fibers, averaged over 5 spectra and normalized relative to the 80$\mu$L data point. The red error bar at the 1~$\mu$L point represents the background noise and corresponding uncertainty.  (c) Relative variations in integrated PL irradiance as a function of illumination time. The Illumination source was briefly removed after 7 days to check for any recovery of the PL signal. (d) Residuals of fit from (b) indicating a maximum 2\% deviation from linear behavior.} \label{fig:pl}
\end{figure}

\subsection{Nanocomposite quantum yield}
\begin{samepage}
To characterize any potential loss of brightness upon incorporating cQDs within an optical fiber, the QY has been measured as a function of cQD concentration in hexane, preform, and fiber. For the latter, an indirect measurement was performed using preform samples which were subjected to the same heating treatment used to draw fibers. For each concentration of cQDs in preform and fiber, the QY is presented relative to its value for the cQD solution at the same concentration (Figure 6). This latter value was measured to be 0.20, as reported for CdSe/CdS cQDs of comparable size in previous studies at room temperature \cite{vela, kang}.\par
\end{samepage}
Following the polymerization stage, the data displayed in Figure 6 show a drop in QY of around 35\% from the normalised value in hexane, with no dependency in regards to the cQD concentration. In contrast, the QY values observed in preform samples before and after heating lie within measurement uncertainties  indicating the fiber drawing process has little to no effect on the QY of all nanocomposites. Given a significant change in QY occurs during polymerization which is accompanied by the PL redshift discussed in section 3.5, cQD surface related mechanisms, rather than inter-dot energy transfer, are thus further validated as causative factors for altering the PL emission dynamics. This is in agreement with conclusions of the Brovelli and Klimov groups implying the CdS shell thickness can be adjusted to minimize the QY drop in nanocomposites\cite{Meinardi}.

\begin{figure}[htpb]
\centering
\includegraphics[height=8cm]{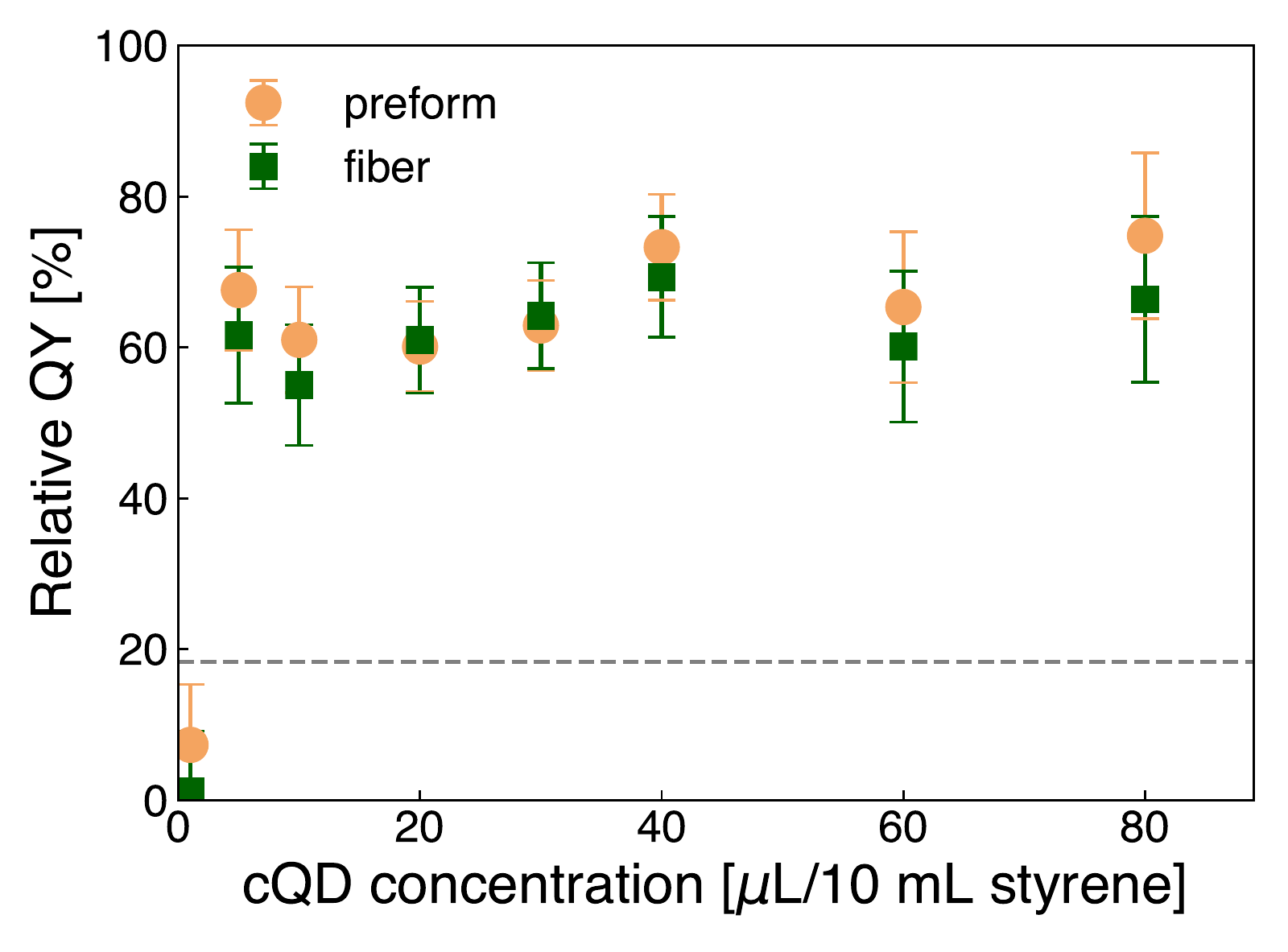} 
\caption{QY of nanocomposites exposed to heat during the fiber drawing process (fiber green squares) and compared to the original  preforms (yellow circles) as a function of cQD concentration, taken relative to the QY of the same concentration of cQDs in hexane. The 1$\mu$L sample is an outlier here since its value is below the measurement noise floor (dashed grey line).  Therefore, no significant change in QY is observed in all cases.} \label{fig:qy}
\end{figure}

\subsection{Refractive index of nanocomposite preforms}

Finally, accurate knowledge of the nanocomposite refractive index will be required to design the waveguiding properties of new optical fibers. Measured values of the refractive index, $n$, are shown in Figure~7 for all nanocomposite preforms, along with blank polystyrene as the reference value of $n=$1.59. No change in the refractive index was observed within 1$\sigma$ standard deviation of polystyrene for the range of concentrations studied. Although a corresponding increase in refractive index could be expected when doping a polymer matrix with a higher polarizability material, it appears here that, despite the cQD dopant concentration being high enough to provide sufficient brightness for many practical applications, it does not impact the refractive index at the measured wavelength. Indeed, when considering the highest cQD concentration of 10$^{7}$~cQDs/cm$^{3}$ for the 80~$\mu$L sample, and spherically shaped cQDs with a 15~nm average diameter, the cQD volume fraction of only 0.004$\%$ is an almost negligible part of the polymer host. Consequently in Figure~7, preform density measurements confirm a nearly constant value within 1$\sigma$ standard deviation for all samples, independently of their dopant concentration.

\begin{figure}[htpb]
\centering
\includegraphics[height=8cm]{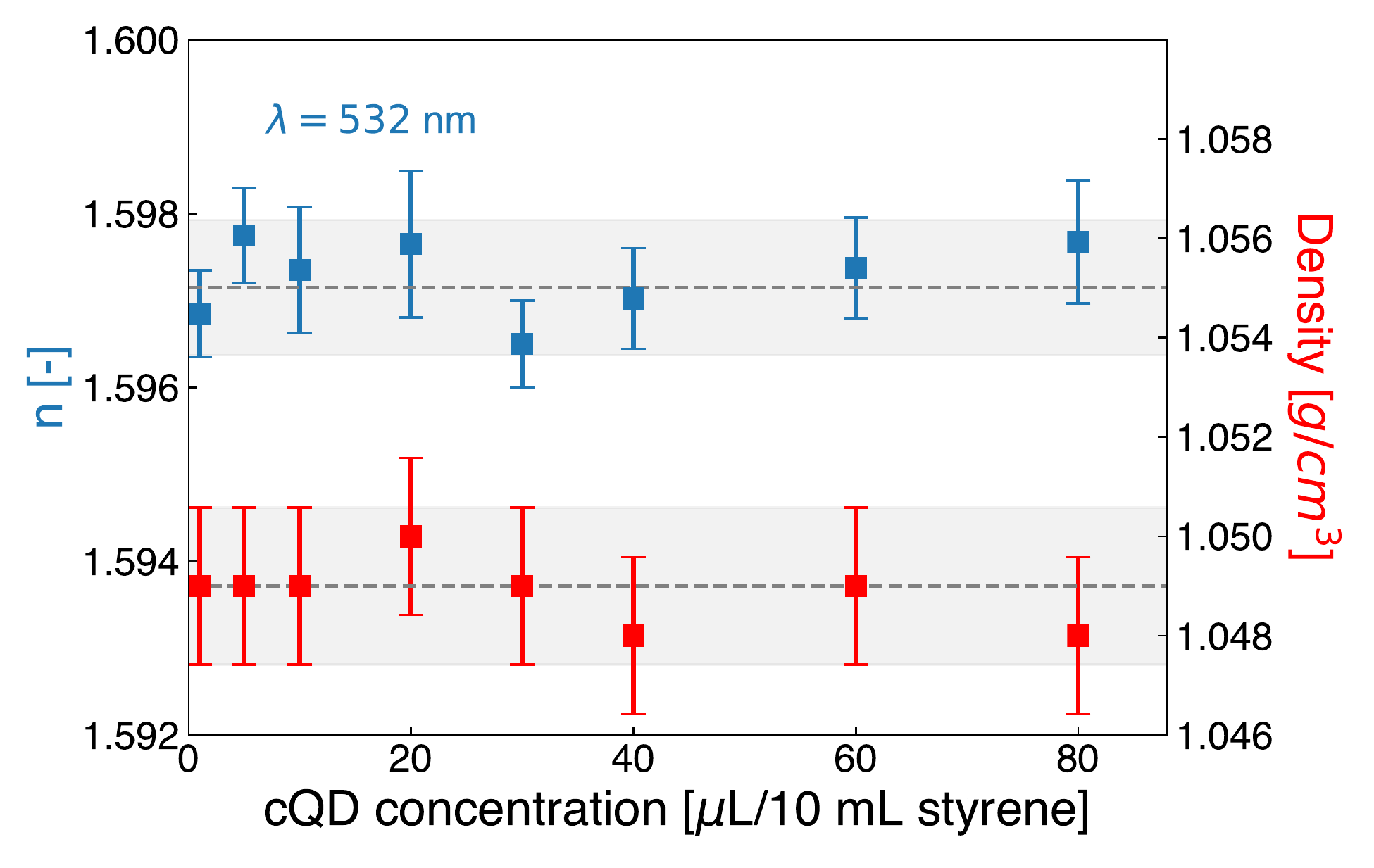} 
\caption{Refractive index (blue) and density (red) of cQD-doped preforms, both being constant within 1$\sigma$ (shaded area, error bars) of blank polystyrene (dashed line).} \label{fig:ri}
\end{figure}

\subsection{Conclusion}

In this paper, we have reported a plastic optical fiber truly doped with core-shell colloidal semiconductor quantum dots. The homogeneous integration of cQDs within the optical fiber core, as demonstrated here, enables the creation of compact and highly customizable light generating fiber-based devices with relevant applications in a range of fields including medicine and biology.\\   
Using the fabrication method described here, cQDs were found to be well dispersed and largely unaggregated throughout the core polymer matrix, an essential requirement in minimizing fiber scattering losses. Degradation of the cQD surface passivation during polymerization affected mostly the QY, but the PL emission irradiance was remarkably stable, varying by less than 5$\%$ over 12 days. The loading fraction of cQDs in the fiber core has been varied to investigate the effects on the optical and material properties of the nanocomposite fibers. Most remained unchanged thanks to the loading fraction kept below 0.004\% that, given the large cQD absorption cross-section, still yields sufficient brightness for most applications. These results represent the first stage in the development of a new class of highly customizable light-generating optical fibers.

\section{Experimental methods}

\noindent{\textbf{Chemicals:}} Styrene (Alfa Aesar\texttrademark, 99$\%$ stabilized with 10-15 ppm 4-tert-butylcatechol), aluminium oxide powder 60 mesh (Alfa Aesar\texttrademark, Brockmann Grade I basic, 58 angstroms), hexanes (certified ACS), and isopropanol (certified ACS+) were purchased from Fisher Scientific. Benzoyl peroxide (Luperox\textregistered~A98, reagent grade $\geqslant$98$\%$), oleylamine (technical grade 70$\%$), 1-octadecene (technical grade 90$\%$), oleic acid (technical grade 90$\%$), trioctylphosphine oxide (TOPO, technical grade 90$\%$), trioctylphosphine (TOP, technical grade 90$\%$), selenium powder 100 mesh ($\geqslant$99.5$\%$ trace metals basis), rhodamine 6G (dye content $\sim$95$\%$) and fluorescein (free acid, dye content $\sim$95$\%$) were purchased from MilliporeSigma. Sodium hydroxide (ACS reagent grade $\geqslant$97$\%$) was purchased from Anachemia VWR. Sulfur powder 100 mesh (grade 99$\%$) and cadmium oxide (grade 99.5$\%$) were purchased from Laboratoire MAT.\\

\noindent{\textbf{cQD synthesis and dispersion:}} The cQDs prepared for this study consist of a CdSe core ($\text{diameter}\sim$3.2 nm) surrounded by a CdS shell, which were synthesized following the methods by Nasilowski \textit{et al.}\cite{nasilowski} Afterwards, the cQDs are purified using several precipitation and centrifugation cycles using isopropanol and hexane as the cQD non-solvent and solvent respectively, then re-dispersed in 10 mL of hexane to obtain a 6 $\mu$M concentration.\\

\noindent{\textbf{Preform fabrication:}} To prepare 10-cm preforms, commercial styrene was first filtered through a column containing Al$_{2}$O$_{3}$ particles to remove the polymerization inhibitor. A fixed volume of the cQD solution (Table 1) was further purified by precipitation and centrifugation with isopropanol then re-dispersed progressively in styrene and transferred to a glass test tube. To initiate the radical polymerization, 60~mg of initiator Luperox\textregistered~A98 per 10~mL of purified styrene was added to the test tube. Oxygen was purged from the sample with nitrogen, then the tube was sealed and kept under vacuum in an ultrasonic bath for a few minutes to remove residual oxygen. To prevent aggregation of the cQDs, a preliminary polymerization stage was carried out by placing the samples in a heated ultrasonic bath ($\sim$60$^{\circ}$C) for 7~hours until the nanocomposite mixture had a syrup-like consistency. After further removal of residual oxygen within the glass tubes still in the bath, the samples were placed in an oven at 90$^{\circ}$C for 48~hours until fully polymerized.\\

\noindent{\textbf{Optical fiber drawing:}} To fabricate the fibers, a 14.3-mm diameter core preform was matched with the thinnest PMMA cladding tube commercially available, namely a thickness of 1.5~mm with an inner diameter of 15.8~mm, to produce a fiber with a large core/cladding ratio. Both the preforms and PMMA tubes are cleaned carefully with soap and deionised water to remove any dust or impurities which could get trapped at the core/cladding interface during fiber drawing. They are then annealed for at least 24 hours in an oven under vacuum, just below their $T_g$ where deformation is avoided. This helps to remove any residual stress within the material and desorb water$/$solvents which could induce defects in the fiber during drawing. To draw the fiber, the core preform is placed within the PMMA tube and centred within the drawing furnace. In order to prevent oxidation during heating and ensure the cladding is sealed around the core material, the gap between the core and cladding materials is lightly evacuated under vacuum for the entire drawing process. This whole structure is then heated to a temperature sufficiently above $T_{g}$, in this case between 155$^{\circ}$C and 160$^{\circ}$C. During fiber drawing, the furnace temperature, preform feed rate and drum speed were adjusted to achieve the desired outer diameter of $\sim$1000 $\mu$m, see Figure \ref{fig:pretty}.\\

\noindent{\textbf{Differential Scanning calorimetry:}} DSC was performed using a Netzsch DSC Pegasus 404F3. Samples were heated in an aluminium pan from -30$^{\circ}$ to 150$^{\circ}$C at a rate of 20$^{\circ}$C/min. In order to remove any effects related to thermal history or residual stress, three scans were performed on each cQD-doped polymer sample. The T$_{g}$ was evaluated by measuring the onset temperature of the baseline shift in the DSC thermogram of each nanocomposite.\\
\newpage
\noindent{\textbf{TEM imaging:}} Thin slices of cQD-doped preforms and a reference of cQDs in solution dropcasted on a carbon grid (Ted Pella 01824) were imaged with a JEOL JSM-1230 TEM operated at 80 kV. Using a batch processing macro in ImageJ, the micrographs were smoothed and background subtracted using a rolling ball algorithm. Further processing was performed to yield cQD positions, areas ($A$), perimeters ($P$) and Feret’s diameters. Each cQD radius is taken as the average of $P/2\pi$ and $\sqrt{(A/\pi)}$, which was verified smaller against the half of both Feret’s diameters.\\

\noindent{\textbf{Nanocomposite transmission and PL spectroscopy:}} Each nanocomposite was first machined and polished to fit in the same square, with 1~cm propagation path length, as the cuvette used for the reference stock solution of cQDs in hexane. Absorbance spectra were then measured with a UV-VIS spectrophotometer (Agilent Cary 50). Ratios with blank reference measurements are done automatically by the instrument software, the data was converted to a linear scale as needed. PL spectra of all cQD-doped samples were obtained with an Ocean Optics 2000USB spectrometer using a 400~nm UV diode ($\sim$75 $\mu$W peak power, $\sim$4~nm FWHM) as the excitation source.\\

\noindent{\textbf{Fiber transmission spectroscopy and PL:}} The cut-back method was used to measure the attenuation of the optical fibers between 450~nm and 800~nm \cite{hect}. Spectra were acquired using a broadband halogen lamp through a filter avoiding second order diffraction at the detection by a optical spectrum analyser (ANDO AQ6135A). Variations in fiber end-face quality affect the core light coupling efficiency and subsequent power transmission. Hence, this was assessed by repeatedly and carefully polishing the fiber end-face flat at each measurement length. To ensure that only light which propagated along the whole fiber length in the core was measured, a black carbon-based paint (Ted Pella DAG-T-502) was applied along the cladding at each end of the fiber near the collection and launch points, thus absorbing light propagating in modes within the PMMA cladding. To roughly compare with cQD absorption in solution and preforms, the nanocomposite fiber transmission spectra were also normalized by a blank fiber reference after scaling for a 1~cm segment.\\
To measure the PL of each cQD-doped fiber with little influence from the polymer matrix, the collection end of a 5-mm long segment was polished, and the opposite end blackened to prevent back-reflected light reaching the detector. Each fiber tip was butt-coupled to a PMMA light guide to record the PL spectrum with an Ocean Optics 2000USB spectrometer upon sideways 90$^{\circ}$ excitation with the same 400~nm UV diode used for nanocomposite characterization. To estimate coupling uncertainties, the cQD-doped fiber was removed and recoupled to the light guide ten times, and its PL spectrum recorded for every re-connection then averaged. The total light irradiance output of each fiber is calculated by integrating its PL emission spectrum over the cQD emission peak from 580~nm to 720~nm. To assess the emission stability of our cQD-doped fibers, the same experimental methods to monitor PL over time were followed, but using instead a 15-cm long piece of fiber with 20~$\mu$L cQD doping.\\

\noindent{\textbf{Quantum yield measurements:}} Samples were prepared and absorbance spectra were acquired with the same methods as previously used for nanocomposites above, while the PL excited using a 488~nm argon laser was measured with the same Ocean Optics spectrometer and integrated to obtain the corresponding PL intensity. For the reference cQDs in hexane, their QY was cross-validated using the standards rhodamine~6G in distilled water and fluorescein in 0.1~M NaOH, with measured values of 0.99 and 0.90 respectively. Uncertainties were estimated from the signal-to-noise ratio in the spectra, except for two samples with the highest concentrations of cQDs where scattering and the inner filter effect were taken into account.\\

\noindent{\textbf{Refractometry and densimetry:}} Nanocomposite samples were cut in disks of about 1~cm in diameter by 0.6~cm in height, then polished to obtain optically clear surfaces in order to measure their refractive index at an energy of 2.33~eV ($\lambda=$532~nm) above the semiconductor band gap. The data acquisition was repeated 20 times on different surfaces for each sample with a Metricon 2010/M prism coupler instrument, from which a 1$\sigma$ uncertainty was estimated for each distribution. The densities for all these samples, but the 20~$\mu$L nanocomposite, were recorded four times by buoyancy measurements in distilled water on a Mettler-Toledo XSE204 analytical balance. The density of the 20~$\mu$L cQD-doped preform was measured 35 times instead to estimate the uncertainties.

\subsection{Supporting Information}
The following files are available free of charge.\\
Size histogram of colloidal quantum dots (cQDs) in hexane as measured from transmission electron microscopy (TEM) data of a 1/100 dilution from the cQD stock solution in hexane, fitted to a log-normal distribution (PDF); TEM micrographs of cQDs dispersed in polystyrene preforms for concentrations between 80$~\mu$L and 1~$\mu$L per 10~mL styrene (PDF); TEM micrograph showing a highly aggregated cluster of cQDs in a 200~$\mu$L doped nanocomposite fabricated without pre-polymerization in a hot ultrasonic bath and a comparison of the normalized photoluminescence (PL) spectra of 40~$\mu$L cQD-doped preforms produced with and without a pre-polymerization stage (PDF);
Histograms of the inter-cQD distances between nearest neighbours compared to the theoretical cumulative distributions for a random Bernoulli process governing the spatial dispersion of cQDs in the polymer matrix (PDF).

\begin{acknowledgement}
The authors thank Patrick Larochelle for technical assistance, Mourad Roudjane, Kishore Kumar and Yannick Ledemi for fruitful discussions as well as Louis-Philippe Dallaire and Justine Giroux for preliminary polymerizations and illustrations in the latter case. Kind suggestions from Mathieu Boivin, Marc-Antoine Langevin and Evelyne Brown Dussault are recognized. Financial support from the Fonds de Recherche du Québec – Nature et technologies (FRQNT), the Natural Sciences and Engineering Research Council (NSERC) and the Canadian Excellence Research Chair program (CERC) in Photonics Innovations is gratefully acknowledged.
\end{acknowledgement}

% \section{Table of contents}
% Nanocomposite optical fibers with customizable properties are realised through polymerization directly embedding colloidal quantum dots for truly homogeneous doping of the polystyrene core matrix. These hybrid, large area active fibers display a bright and stable luminescence over days of excitation, making them relevant for a wide range of smart applications.
% \begin{figure}[htpb]
% \centering
% \includegraphics[height=5cm]{TOCimage.jpg} 
% \caption{}
% \label{fig:TOC}
% \end{figure}

%\bibliographystyle{angew}
\bibliography{cQDbib_vCA}
%\printbibliography

\newpage
\section{Supporting Information}
\subsection{cQD-Doped Optical Fibers Drawn from Polymer Nanocomposites}
Carly A. Whittaker$^{1,2}$, Arthur Perret$^1$, Charles Fortier$^1$, Olivier-Michel Tardif$^{~1}$, Sébastien A. Lamarre$^{1}$, Steeve Morency$^{1}$, Dominic Larivière$^{3}$, Luc Beaulieu$^{2}$, Younès Messaddeq$^{1}$, Claudine Nì. Allen$^1$

\begin{center}
\noindent {\small 1. Centre d’optique, photonique et laser (COPL), Département de physique, de génie physique et d’optique, Université Laval, Québec G1V 0A6, Canada}\\
\noindent {\small 2. Département de physique, de génie physique et d’optique, Centre de recherche sur le cancer, CHU de Québec, Université Laval, Québec G1V 0A6, Canada}\\
\noindent {\small 3. Laboratoire de radioécologie, Département de chimie, Université Laval, Québec G1V 0A6, Canada}\\
\end{center}

\subsection{Supplementary figures}

\setcounter{figure}{0} 
\begin{figure}[htpb]
\centering
\includegraphics[height=8cm]{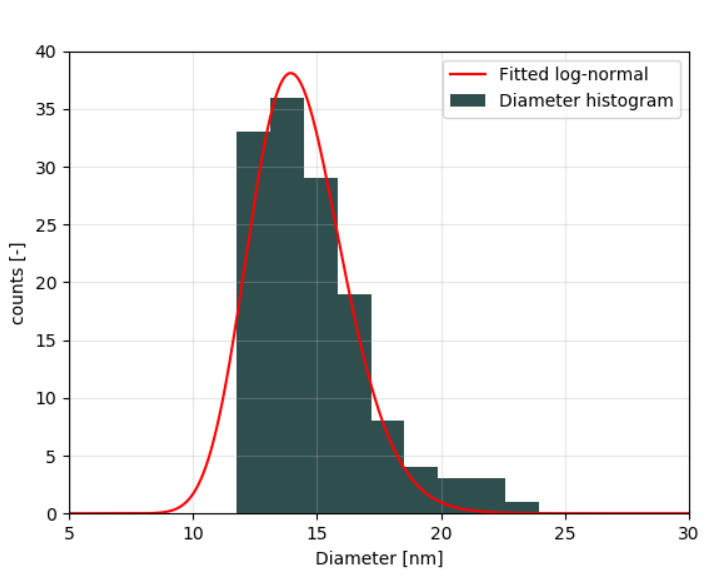} 
\caption{Size histogram of colloidal quantum dots (cQDs) in hexane as measured from transmission electron microscopy (TEM) data of a 1/100 dilution from the cQD stock solution in hexane, fitted to a log-normal distribution (red line). This distribution has a mean diameter of 14.1~nm and a variance of 3.6~nm$^{2}$.} \label{fig:sizesupp}

\end{figure}

\begin{figure}[htpb]
\centering
\includegraphics[height=8cm]{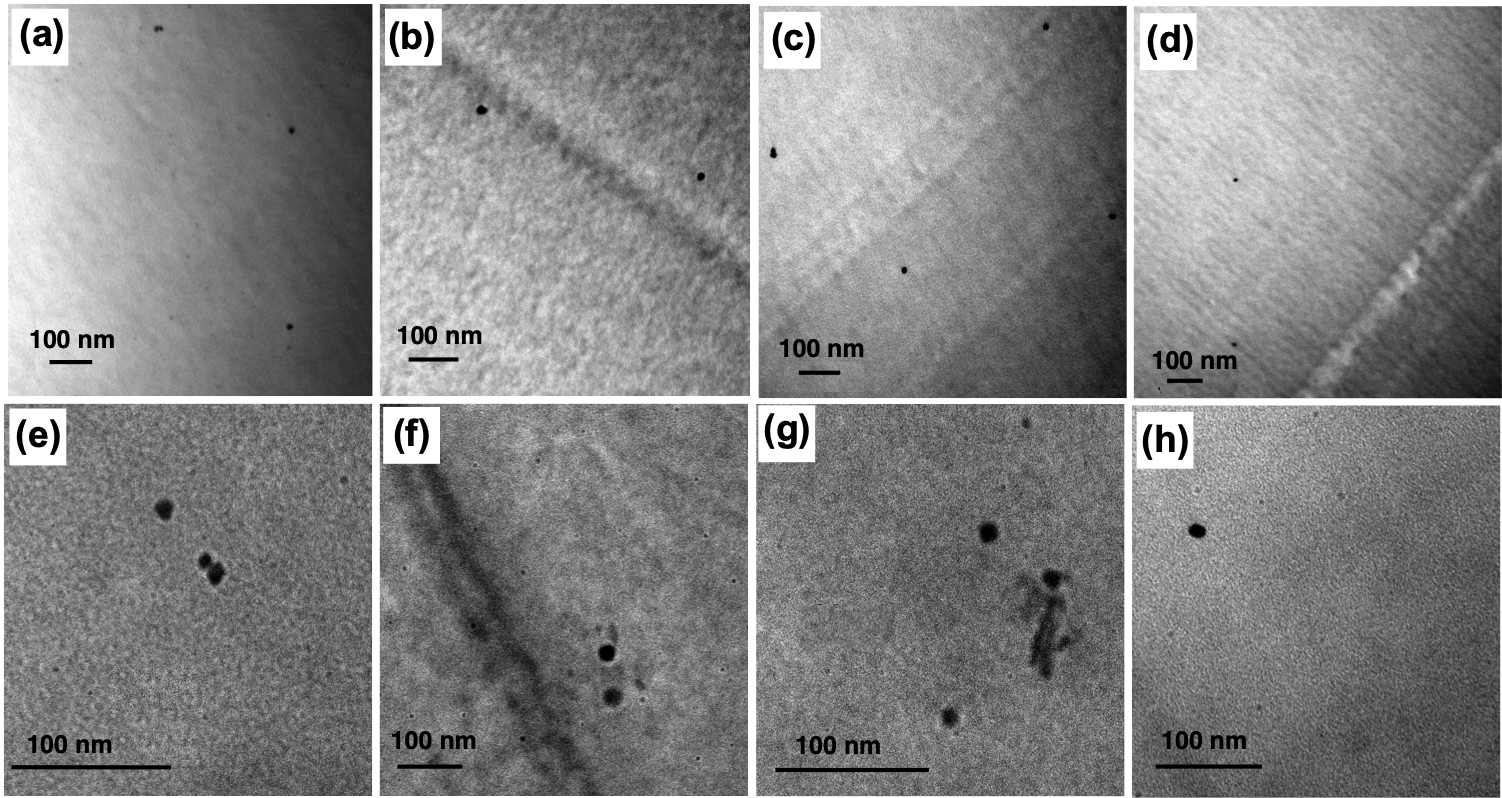} 
\caption{TEM micrographs of cQDs dispersed in polystyrene preforms for concentrations between (a) 80~$\mu$L to (h) 1~$\mu$L per 10~mL styrene. A small cQD pair cluster is seen in~(e).} \label{fig:temsupp}
\end{figure}

  \begin{figure}[htpb]
\centering
\includegraphics[height=8cm]{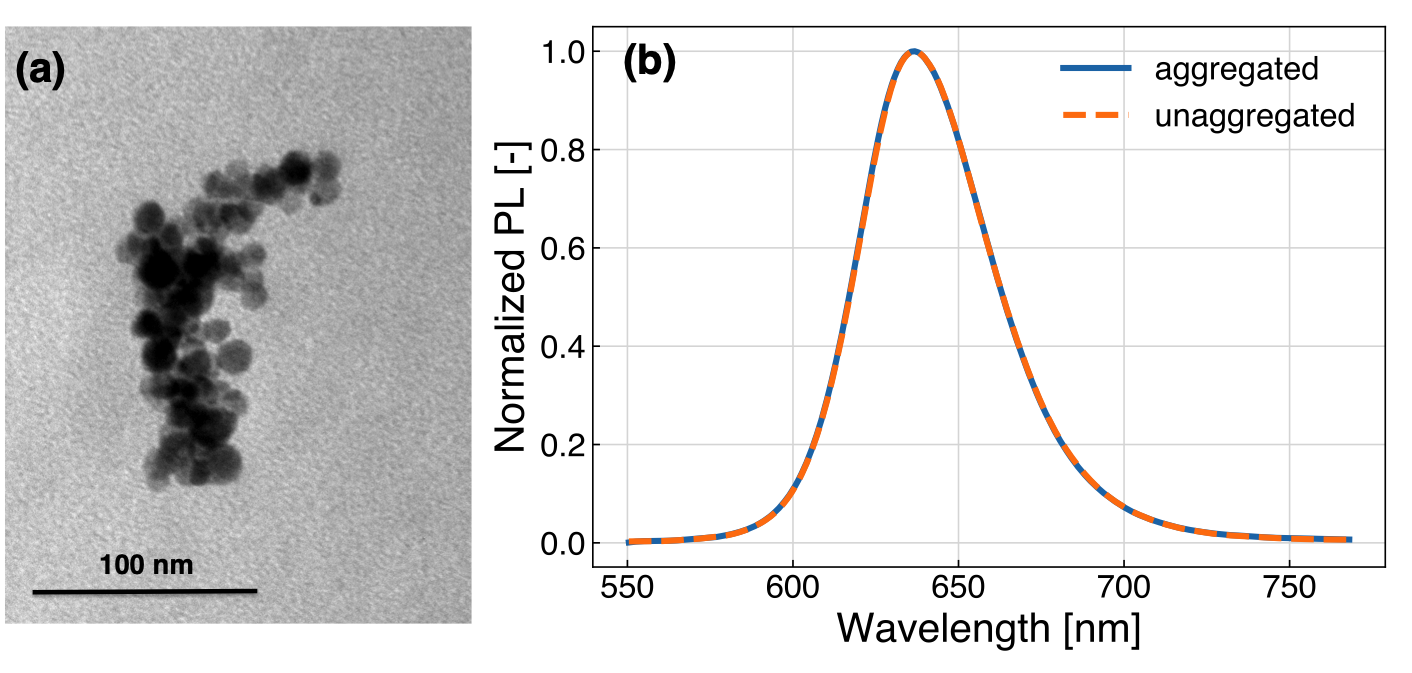} 
\caption{TEM micrograph showing a highly aggregated cluster of cQDs in a 200~$\mu$L doped nanocomposite fabricated without pre-polymerization in a hot ultrasonic bath. (b) Comparison of the normalized photoluminescence (PL) spectra of 40~$\mu$L cQD-doped preforms produced with (orange dashes) and without (blue line) a pre-polymerization stage. The absence of a redshift in presence of agglomerated cQD clusters for the latter sample refutes this aggregation as a mechanism via Förster resonance energy transfer (FRET) to explain the PL change of cQDs incorporated in nanocomposites.} \label{fig:aggsupp}
\end{figure}

\begin{figure}[htpb]
\centering
\includegraphics[height=8cm]{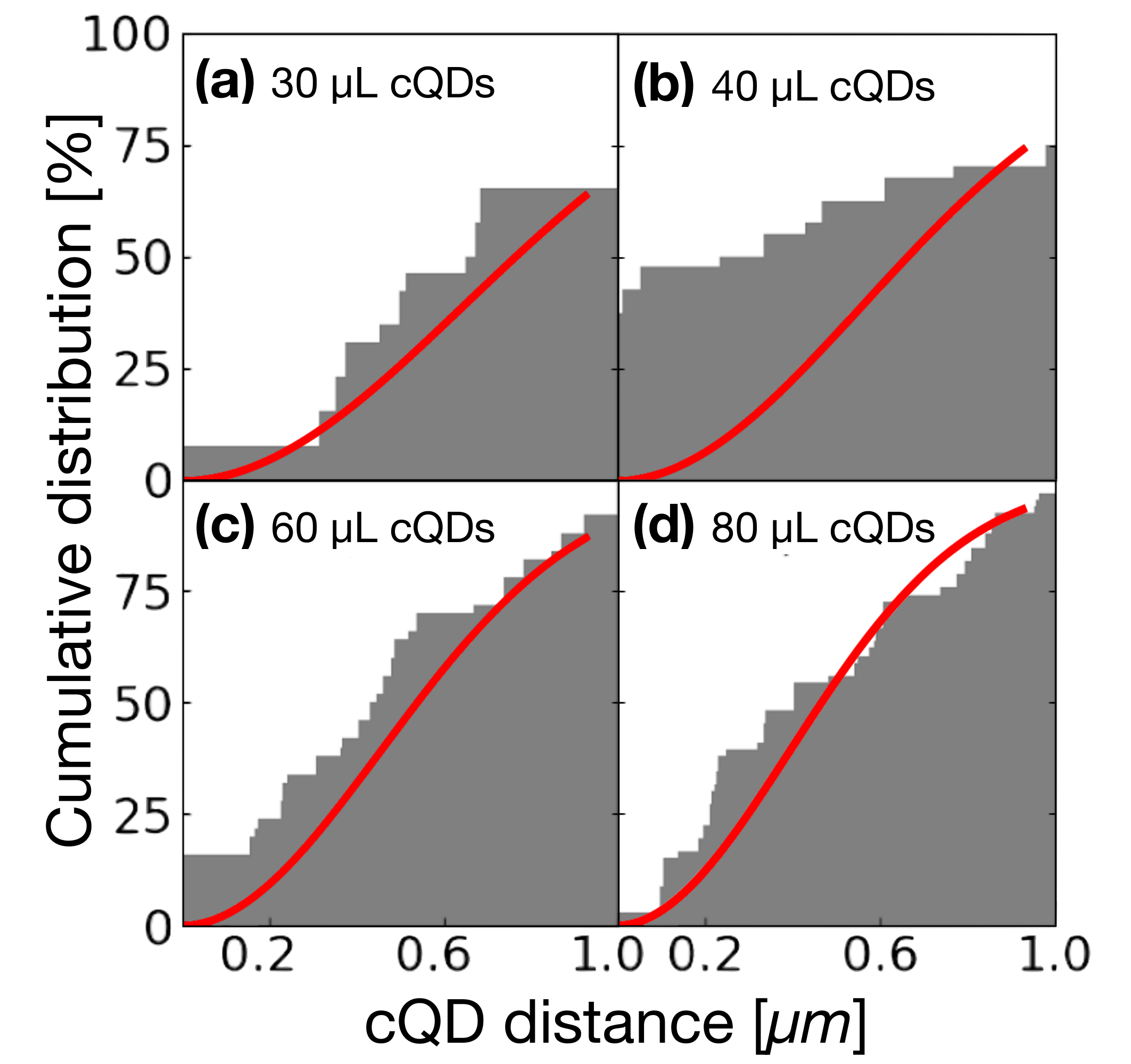} 
\caption{Histograms (shaded in grey) of the inter-cQD distances between nearest neighbours as presented in Figure 2 of the main text, but now compared to the theoretical cumulative distributions (red lines) for a random Bernoulli process governing the spatial dispersion of cQDs in the polymer matrix.} \label{fig:cQDs}
\end{figure}

\end{document}